\documentclass[aps,prl,superscriptaddress,twocolumn]{revtex4-1}
\usepackage[dvips]{graphicx}
\usepackage{dcolumn}
\usepackage{bm}
\usepackage{amsmath,amsthm,amssymb}
\usepackage{epstopdf}
\usepackage{color}

\newcommand{\ket}[1]{\left|#1\right>}

\begin{document}

\title{Correlated Exciton Transport in Rydberg-Dressed-Atom Spin Chains}
\author{H. Schempp}
\author{G. G\"unter}\affiliation{Physikalisches Institut, Universit\"at Heidelberg, Im Neuenheimerfeld 226, 69120 Heidelberg, Germany.}
\author{S. W\"uster}\affiliation{Max Planck Institute for the Physics of Complex Systems, N\"othnitzer Strasse 38, 01187 Dresden, Germany.}
\author{M. Weidem\"uller}\affiliation{Physikalisches Institut, Universit\"at Heidelberg, Im Neuenheimerfeld 226, 69120 Heidelberg, Germany.}\affiliation{Synergetic Innovation Center of Quantum Information and Quantum Physics, and Hefei National Laboratory for Physical Sciences at Microscale, University of Science and Technology of China, Hefei, Anhui 230026, China.}
\author{S. Whitlock}\email{whitlock@physi.uni-heidelberg.de}
\affiliation{Physikalisches Institut, Universit\"at Heidelberg, Im Neuenheimerfeld 226, 69120 Heidelberg, Germany.}
\date{\today}

\begin{abstract}
We investigate the transport of excitations through a chain of atoms with non-local dissipation introduced through coupling to additional short-lived states. The system is described by an effective spin-1/2 model where the ratio of the exchange interaction strength to the reservoir coupling strength determines the type of transport, including coherent exciton motion, incoherent hopping and a regime in which an emergent length scale leads to a preferred hopping distance far beyond nearest neighbors. For multiple impurities, the dissipation gives rise to strong nearest-neighbor correlations and entanglement. These results highlight the importance of non-trivial dissipation, correlations and many-body effects in recent experiments on the dipole-mediated transport of Rydberg excitations.
\end{abstract}

\maketitle

The transport of energy, charge, or spin is of fundamental importance in diverse settings, ranging from the operation of nanoelectronic and spintronic devices~\cite{Joachim2000,Nitzan2003,Zutic2004,Catalan2012}, to the dynamics of electron-hole pairs in organic semiconductors~\cite{Akselrod2014,Menke2014} and in natural processes such as photosynthesis~\cite{Amerongen2000,Scholes2011,Collini2013}. In these systems different elementary excitations and basic transport mechanisms can give rise to very different behaviour including coherent exciton motion, thermally activated diffusion or even collective fluid-like dynamics~\cite{Amo2009}. However, understanding or exploiting these differences (e.g.~for applications in photovoltaic devices) is extremely challenging, as they depend on the complex interplay between quantum statistics, coherence, confinement, disorder, and the nature of the interactions between the constituent particles. A key question is how dissipation, through the coupling to reservoirs, leads to a cross-over between coherent and incoherent motion. While dissipation is usually assumed to destroy coherence, it is becoming evident that certain dissipative processes, noise or specially structured environments may even enhance coherence or enable new modes of transport, both in engineered quantum systems and in natural ones~\cite{Plenio2008,Caruso2009,Scholes2011}.

 \begin{figure}
  \begin{center}
  \hspace*{-0.1cm}
      \includegraphics[width=0.95\columnwidth]{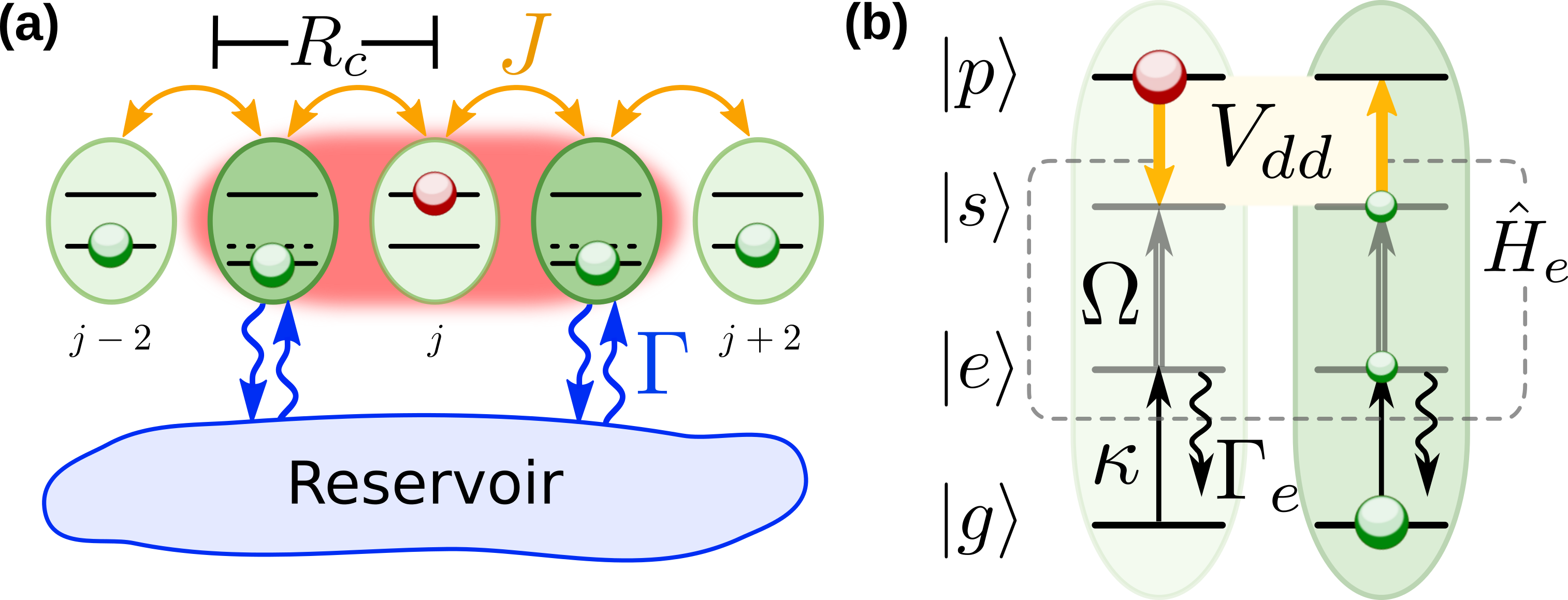}
      \caption{(Color online) Transport of excitations (red spheres) through a chain of optically-dressed-atoms with site index $j$ (green spheres). (a) Excitonic motion arises through coherent long range exchange interactions $J$ and through dissipative processes mediated by a tailored reservoir coupling. (b) Each site is represented by an atom which can be in either the impurity Rydberg state $\ket{p}$ or the auxiliary states $\ket{g}$, $\ket{e}$ or $\ket{s}$. Dipolar interactions result in exchange between $\ket{s}$ and $\ket{p}$ Rydberg states. States $\ket{g}\leftrightarrow \ket{e}$ and $\ket{e}\leftrightarrow \ket{s}$ are coupled with Rabi frequencies $\kappa$ and $\Omega$ respectively. The state $\ket{e}$ spontaneously decays to the $\ket{g}$ state with the rate $\Gamma_e$. The state space $\hat H_e$ comprising $\ket{e}$ or $\ket{s}$ excitations can be adiabatically eliminated resulting in an effective spin-1/2 model for the impurity and dressed states. 
 }
    \label{fig:fig1}
  \end{center}
\end{figure}

Here we analyse the exciton-like motion of individual excitations through a network of dipole interacting quantum systems (i.e.~atoms, molecules or quantum dots) in the presence of a specially engineered reservoir (Fig.~\ref{fig:fig1}a). Each subsystem is assumed to be coupled to one or more additional short-lived states. The populations of these states are determined by the coupling fields and by their proximity to any excitations, which provides a handle to introduce and control new types of dissipation in the system. By eliminating the short-lived states we show that the resulting system is characterised by coherent hopping and non-local dissipative terms which lead to correlated exciton-motion. Simulations of the single excitation and the two excitation dynamics demonstrate this has a dramatic effect on the transport properties, leading for example to the emergence of a new length scale for hopping and strong spatial correlations even at steady state.

As a physically realizable system we propose a chain of ultracold atoms in which Rydberg states with large transition electric dipole moments are optically coupled to low-lying electronic states, which provides a clean synthetic system to investigate energy transfer dynamics and transport including controllable interactions and dissipation~\cite{Mourachko1998,Anderson1998,Anderson2002,Westermann2006,Ditzhuijzen2008,Mulken2007,Wuster2010,Wuster2011,Scholak2011,Bettelli2013,Robicheaux2014,Scholak2014}. 
Recently the exciton-like migration of an ensemble of Rydberg impurities in a bulk system was observed using electromagnetically-induced-transparency (EIT) of a background atomic gas as an amplifier~\cite{Gunter2013}. Remarkably, the figure-of-merit for exciton migration, the diffusion length $L_d=\sqrt{D\tau}\approx 50-100~\mu$m (with $D$ the diffusion coefficient and $\tau$ the typical lifetime of the Rydberg state) was an order of magnitude larger than in even the purest organic semiconductors~\cite{Najafov2010}, and can be tuned by the probing light fields which act as a controllable environment. In another experiment the coherent hopping of a single excitation amongst three equidistantly spaced Rydberg atoms was observed~\cite{Barredo2015}. Future experiments will be able to probe the coupled coherent and incoherent motion with high spatial and temporal resolution, while the strength and nature of the dipole coupling, the degree of disorder, excitation density and the role of the environment can all be controlled, allowing unprecedented opportunities to investigate the fundamental processes at play. 

Consider a system of ultracold atoms initially prepared in either a Rydberg state with orbital angular momentum $l=1$ (impurity state $\ket{p}$) or the electronic ground state $\ket{g}$. The $\ket{g}$ state is weakly optically coupled via an EIT resonance to a short-lived excited state $\ket{e}$ and an $l=0$ Rydberg state $\ket{s}$ (Fig.~\ref{fig:fig1}b). The $\ket{g}\rightarrow\ket{e}$ probe transition is driven with Rabi frequency $\kappa$ and the $\ket{e}\rightarrow\ket{s}$ coupling transition is driven with Rabi frequency $\Omega$ with $\kappa \ll \Omega$. The state $\ket{e}$ spontaneously decays to $\ket{g}$ with a rate $\Gamma_e$, while the other states are assumed to be stable on the time scale of the dynamics. Migration of the $\ket{p}$ excitations among the dressed ground state atoms occurs via the $\ket{s}$ state admixture (by $|s_j\rangle|p_k\rangle\leftrightarrows |p_j\rangle|s_k\rangle$ exchange, with $j$ and $k$ the site indices)~\cite{Wuster2011}, while the $\ket{e}$ state admixture introduces a controllable environment~\cite{Schoenleber2015}, both of which are influenced by the competition between the EIT laser fields and the Rydberg-Rydberg interactions.

The Hamiltonian describing the system is given by ($\hbar=1$)
 \begin{equation}
 \hat{H}=\hat{H}_{0}+\sum_{j\neq k}V_{dd}^{(j,k)}|s_j\rangle|p_k\rangle \langle p_j|\langle s_k|
 \end{equation}
 where $\hat{H}_{0}=\frac{1}{2}\sum_j \Omega |s_j\rangle\langle e_j|+\kappa |e_j\rangle\langle g_j| +\mathrm{h.c.}$ accounts for the single-atom laser couplings and $V_{dd}^{(j,k)}=\nobreak C_m/|x_j-\nobreak x_k|^m$. Here we mainly consider dipolar interactions corresponding to $m=3$, however $m=6$ can also be realised using nonresonant van der Waals interactions between Rydberg states~\cite{Comparat2010,vanBijnen2014,Glaetzle2014}. For simplicity the interaction coefficient $C_m$ is assumed to be independent of $j$ and $k$.  Additionally, spontaneous decay of the intermediate excited states is included through Lindblad terms leading to the master equation for the density matrix $\rho$
  \begin{equation}\label{eq:mastereq}
 \dot{\rho}=-i [\hat{H},\rho]+\mathcal{L}[\rho]
 \end{equation}
 where $\mathcal{L}[\rho]=\sum_j \hat{L}_j\rho \hat{L}^\dag_j-\frac{1}{2}\bigl(\hat{L}^\dag_j\hat{L}_j\rho+\rho\hat{L}^\dag_j\hat{L}_j\bigr )$ and each of the Lindblad operators represents a single decay channel ($\hat{L}_j=\sqrt{\Gamma_{e}}|g_j\rangle\langle e_j|$).  
 
Simulating the open-system dynamics for more than approximately five four-level atoms is beyond what can be readily performed using exact numerical methods (e.g. Monte Carlo wave-function techniques). However, in the weak probe regime most relevant to experiments ($\kappa \ll \Omega,\Gamma_e$) the populations of the $|e\rangle$ and $|s\rangle$ states are always small. Furthermore the dynamics between these states due to the laser coupling is fast compared to the hopping dynamics. Therefore the many-body states which include $\ket{e}$ or $\ket{s}$ can be adiabatically eliminated~\cite{Petrosyan2011,Carmele2014,Marcuzzi2014}. 

For this we use the effective operator approach~\cite{Reiter2012}, which allows for a simple interpretation of the dominant processes in terms of coherent and incoherent coupling rates. First the state space is separated into a slowly evolving `ground-state' subspace involving only $\ket{g},\ket{p}$ states and a rapidly evolving `excited-state' subspace including $\ket{e},\ket{s}$ states (Fig.~\ref{fig:fig1}b). The ground state Hamiltonian contains no direct couplings $\hat H_g=0$, while the coupling laser and the dipole-dipole interactions enter the Hamiltonian for the excited state manifold $\hat H_e$, and the probe laser weakly couples ground and excited subspaces through $\hat V_+=\hat V_-^\dagger$. The master equation governing the evolution of the ground states is then defined through the operators $\hat H^\mathrm{eff}=-V_-\mathrm{Re}[\text{x}]+\hat H_g$ and $\hat L^\mathrm{eff}_j=\hat L_j \text{x}$, where $\text{x}$ is the solution to the matrix equation $\hat H_{NH}\text{x}=\hat V_+$ with the non-Hermitian operator $\hat H_{NH}=\hat H_e-(i/2)\sum_j \hat L_j^\dagger\hat L_j$. To simulate the time evolution of the system for a given set of parameters we first numerically obtain $\text{x}$ in order to define an effective master equation [as in Eq.~\eqref{eq:mastereq} but using the operators $\hat H^\mathrm{eff}$ and $\hat L_j^\mathrm{eff}$] which can then be solved in the usual fashion.

\emph{Single exciton dynamics:-} In the special case of a single $\ket{p}$ excitation and by neglecting beyond second-order interactions (corresponding to couplings between states involving more than one dressed-atom), simple analytic expressions for the effective operators can be found. The resulting $N$-site effective master equation can be written
\begin{eqnarray}\label{eq:effectiveoperators}
\hat{H}^\mathrm{eff}&=&\sum_{k>j}J(d_{j,k})\hat{S}^k_+\hat{S}^j_-+h.c.\nonumber \\ 
\hat{L}^\mathrm{eff}_j&=&\sum_{k\neq j}i\sqrt{\gamma(d_{j,k})}\hat{S}^k_+\hat{S}^k_--\sqrt{\Gamma(d_{j,k})}\hat{S}^k_+\hat{S}^j_-
\end{eqnarray}
where $d_{j,k}=|x_j-x_k|$ and we have made use of the spin raising and lowering operators ($\hat{S}_+$ and $\hat{S}_-$ respectively with $\hat{S}_+^ k=\ket{p_k}\langle \tilde g_k|$ where $\ket{\tilde g}$ is the dressed ground state).

The resulting effective operators involve three terms: (1) effective coherent exchange interactions $J=\nobreak\frac{K V{(d_{j,k})}}{2+2V{(d_{j,k})}^2}$, (2) incoherent hopping $\Gamma=\nobreak\frac{K V(d_{j,k})^2}{(1+V(d_{j,k})^2)^2}$ and (3) irreversible dephasing $\gamma=\nobreak\frac{K V(d_{j,k})^4}{(1+V(d_{j,k})^2)^2}$ acting on each site. Here $K=\kappa^2/\Gamma_e$ and $V=(R_c/d_{j,k})^3$ where $R_c=(2\Gamma_e C_3/\Omega^2)^{1/3}$ is the dipole blockade radius~\cite{Gunter2012,Gunter2013}. These two parameters have simple interpretations as the two-level-atom photon scattering rate, and the dipole-dipole interaction energy scaled by the EIT bandwidth respectively. $K^{-1}$ defines a natural timescale for the dynamics whereas $V(d_{j,k})$ is responsible for the competition between coherent exchange, hopping and dephasing. Equations~\eqref{eq:effectiveoperators} can be used to efficiently simulate the open-quantum-system dynamics of a single impurity immersed in a background of hundreds of optically-dressed atoms in arbitrary geometries. It is interesting to note that this effective model is closely related to the widely used Haken-Strobl-Reineker (HSR) model for exciton motion in the presence of noise~\cite{Haken1973,Reineker1973}, including two distinct decoherence mechanisms which originate from the spontaneous decay of the $\ket{e}$ states. However, the incoherent hopping jump operators have an unusual form: $\sum_{k}\alpha_{j,k}\hat S_+^k\hat S_-^j$, describing correlated jumps from states $\ket{p_j}$ to collective states involving many neighbouring sites $k$~\cite{[{A similar effect was also recently noted in }]Marcuzzi2014}. This is in contrast to the original HSR model which ignores the influence of the system on the reservoir and assumes uncorrelated fluctuations e.g.~$\hat L_{j,k}^\text{HSR}=\nobreak\alpha_{j,k}\hat S_+^k\hat S_-^j$. Therefore, the system studied here offers the possibility to use dissipation controlled via EIT resonances to control both the strength and nature of decoherence in these systems which, for example, is an active ingredient of proposals for quantum state preparation by reservoir engineering~\cite{Verstraete2009,Diehl2008,Weimer2010,Weimer2011,BhaktavatsalaRao2013}. 

\begin{figure}
  \begin{center}
  \hspace*{-0.7cm} 
      \includegraphics[width=1\columnwidth]{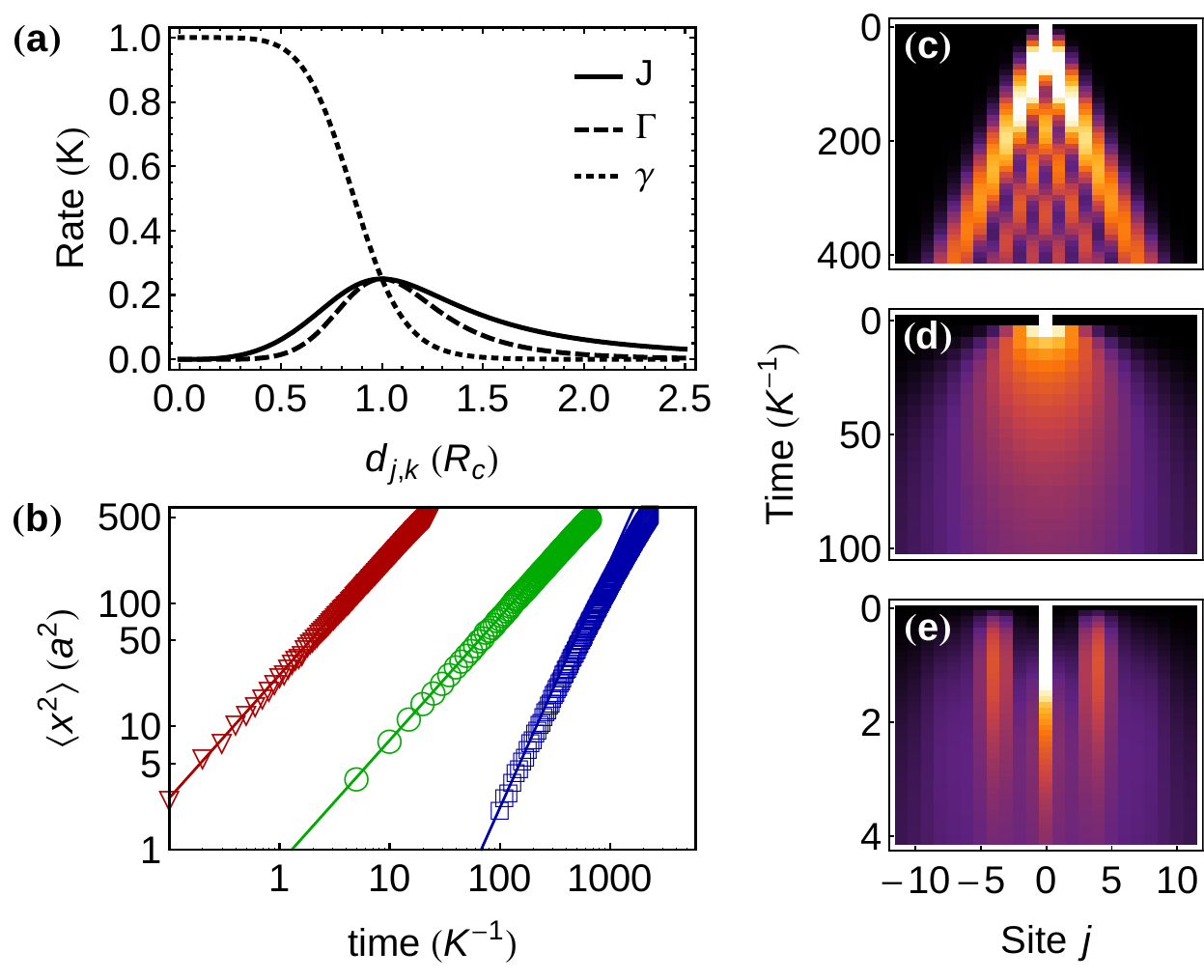}
      \caption{(Color online) (a) Distance dependence of the effective coupling rates $J$ (solid line), $\Gamma$ (dashed line) and $\gamma$ (dotted line) for a single impurity with $m=3$ in units of the two-level atom scattering rate $K=\kappa^2/\Gamma_e$. (b) Mean square displacement of the propagating excitation $\langle x^2\rangle$ for a chain of $N=121$ sites obtained from simulations for three values of the dimensionless interaction strength: $V(a)=R_c^3/a^3=50$ (triangles), $V(a)=1$ (circles) and $V(a)=0.02$ (squares). The solid lines show the approximate analytic scalings in the three regimes described in the text. (c-e) Zoomed in density plots showing the impurity probability distribution at short times for (c) coherent transport with $V(a)=0.02$ (d) diffusive transport with $V(a)=1$ and (e) transport in the blockade regime $V(a)=50$. Bright colors indicate high probability density. 
 }
    \label{fig:fig2}
  \end{center}
\end{figure}

Fig.~\ref{fig:fig2}(a) shows the rates for coherent exchange, hopping and dephasing terms as a function of the relative distance between two sites. For short distances coherent exchange is suppressed while at large distances it is determined by the $d_{j,k}^{-3}$ dependence of the dipolar interaction $J\approx (\kappa^2/\Omega^2)V_{dd}$. The peak exchange rate occurs for $d_{j,k}=R_c$ where $J=K/4$. At this distance however incoherent hopping and irreversible dephasing are equally important. For $d_{j,k}\rightarrow 0$ dephasing dominates, saturating at a rate given by the two-level atom scattering rate $\gamma=K$.

We now turn to the analysis of spin dynamics in this system for the case of atoms arranged in a one-dimensional chain with intersite separation $a$ (as can be produced for example in an optical lattice). We anticipate three main regimes: (1) for $V(a)\ll 1$ coherent exchange dominates over decoherence, (2) for $V(a)\approx 1$ decoherence becomes important leading to classical hopping, (3) for $V(a)\gg 1$ strong dephasing suppresses coherent exchange, leaving incoherent hopping with a characteristic hopping distance comparable to $R_c$. Simulations for a single impurity for $N=121$ sites are shown in Figs.~\ref{fig:fig2}(b-e). The strength of the dipolar interactions is varied through the nearest neighbour Rydberg-Rydberg interaction coefficient $V(a)=R_c^3/a^3$. We calculate the impurity probability distribution $P_j(t)=\mathrm{Tr}(\hat n_j \rho(x_j,t))$ (where $\hat n_j=S^ j_+\hat S^j_-$) on each site for different values of $V(a)$.

 For $V(a)\ll 1$ the dynamics are characteristic of a quantum random walk, with light-cone-like spreading and interference fringes in the spatial density distribution (Fig.~\ref{fig:fig2}c). In this regime coherent dipolar exchange dominates leading to ballistic expansion of the wavepacket. For short times $t\lesssim \Gamma(a)^{-1}$ the mean square displacement (including beyond-nearest neighbor exchange) evolves according to $\langle x^2\rangle/a^2 \approx (\pi^4/180)V(a)^2K^2t^2$ [Fig.~\ref{fig:fig2}(b)]~\cite{Reineker1980}. For intermediate interactions ($V(a)\approx 1$) the coupled coherent and incoherent motion makes the dynamics more difficult to describe, however, from our simulations we find normal diffusion (Fig.~\ref{fig:fig2}d) with $\langle x^2\rangle/a^2 = 2Dt$ where the diffusion coefficient $D\approx [0.262+0.123 V(a)] K$ is found by expanding around $V(a)\approx 1$. For $V(a)\gg 1$ qualitatively new behavior is observed. In this regime there is a new preferred hopping distance given by the Rydberg blockade radius $R_c\gg a$ (Fig.~\ref{fig:fig2}e). This can be understood as the presence of an impurity shifts the $\ket{s}$ states of nearby atoms which suppresses energy transfer. In this regime  the underlying lattice geometry becomes less important, leading for example to a reduced influence of possible disorder in the atomic positions. An expression for the mean square displacement can be found by neglecting coherent exchange and integrating the incoherent hopping rate for all other sites $\langle x^2\rangle/a^2= (\pi/6)V(a)K t$. This scaling differs from the simple picture reported in \cite{Gunter2013}, where continuous observation of the system via EIT was assumed to lead to a Zeno-like slowdown of the dynamics. Instead we demonstrate that the system exhibits a more complex form of dissipation, which allows for rapid transport even in the strongly dissipative limit.

 \begin{figure}
  \begin{center}
    \hspace*{-0.3cm} 
      \includegraphics[width=1\columnwidth]{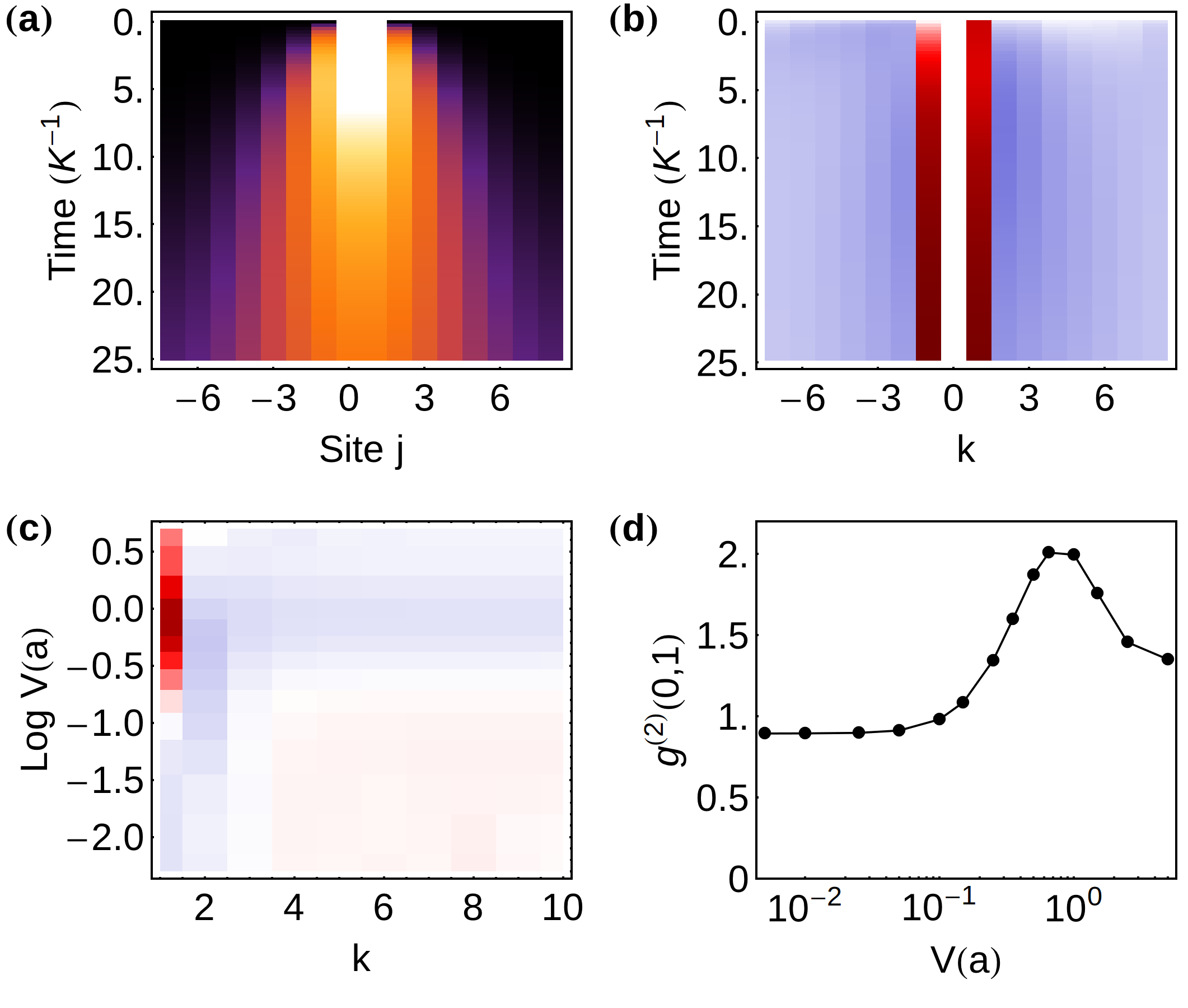}
      \caption{Two exciton dynamics and density-density correlations $g^{(2)}$ for $\Gamma=1$, $\Omega=2$, $K=0.05$, $m=3$ and varying $V(a)$. (a) Probability density as a function of time for $V(a)=1$ showing normal diffusion. (b) Equal time correlation function $g^{(2)}(0,k)$ for the same parameters. The colors correspond to $g^{(2)}>1$ (dark red), $g^{(2)}<1$ (blue) and $g^{(2)}=1$ (white). (c) Equal time density-density correlation function at steady-state as a function of $V(a)$ [same colorscale as (b)].  (d) Nearest neighbour correlation function $g^{(2)}(0,1)$ versus $V(a)$ showing bunching around $V(a) \approx 1$.
 }
    \label{fig:fig3}
  \end{center}
\end{figure}

\emph{Exciton-exciton correlations:-} For times which are long compared to the inverse dissipation rate, each of the different regimes described in Fig.~\ref{fig:fig2} exhibit diffusive behavior and for a finite system the excitation becomes uniformly distributed over the chain (Fig.~\ref{fig:fig3}a). However, even at steady-state ($\dot \rho=0$) the nature of the transport may be revealed through the analysis of higher-order statistical properties. To demonstrate this, and to point out the importance of many-body effects, we present simulations of the effective master equation for two excitations, which could be experimentally prepared, for example, by tuning the excitation laser frequency to match the van der Waals interaction energy between a pair of $p$-excitations at a well defined distance~\cite{Amthor2010,Weber2015}. Time-dependent simulations for $N=16$ sites for intermediate interaction strength are presented in Fig.~\ref{fig:fig3}(a,b) and solutions to the steady-state effective master equation for a chain of $N=20$ sites with varying interaction strength (with periodic boundary conditions) are shown in Fig.~\ref{fig:fig3}(c,d). While the density distribution evolves similarly to the case of a single excitation [Fig.~\ref{fig:fig3}(a)], we observe strong density-density correlations $g^{(2)}(j,k)=\nobreak c\langle \hat n_j \hat n_k\rangle/(\langle \hat n_j \rangle\langle \hat n_k\rangle)$ where the coefficient $c=\nobreak (2N-2)/N$  normalizes to the case of precisely two excitations distributed over the chain in an uncorrelated manner [Fig.~\ref{fig:fig3}(b)]. These correlations persist even at steady-state, long after the memory of the initial state is lost [Fig.~\ref{fig:fig3}(c)]. Fig.~\ref{fig:fig3}(d) shows how the nearest-neighbor correlations depend on $V(a)$, exhibiting strong bunching for $V(a)\approx 1$. For both strong and weak interactions the correlations are suppressed, indicating that they arise as a consequence of the competition between dipole-mediated exchange and distance dependent dissipation provided by the optically-dressed atoms. This dissipation is minimized for neighboring pairs and can be thought of as inducing an effective attraction between the impurities, similar to effects in~\cite{Ates2012,Lemeshko2013}. These correlations vanish if the correlated jump operators of the effective master equation are replaced by operators which couple localized states alone. In the case of short-range interactions (i.e. van der Waals with $m\geq 6$) these correlations become even more pronounced, and for nearest-neighbor interactions the steady-state corresponds to the entangled pure state $\ket{D}=(N-1)^{-1/2}\sum_{j}(-1)^j\ket{\tilde g_1\ldots p_jp_{j+1}\ldots \tilde g_N}$.

The described impurity plus optically-dressed atom system exhibits rapid spin transport controlled by a single parameter $V(a)=2\Gamma_e C_3/(\Omega^2 a^3)$. In addition to the coherent and incoherent hopping regimes recently investigated in a related disordered system~\cite{Schoenleber2015}, we find a blockade dominated regime in which a most-likely hopping distance emerges given by the dipole blockade radius. In the limit of low impurity densities we derive an effective master equation that can be directly applied to current experiments on Rydberg energy transport in optically driven systems, helping to elucidate the interplay between coherent spin exchange and decoherence due to spontaneous decay of the dressed states. Similar physics could arise in other systems where correlated noise~\cite{Chen2010} or exciton-vibrational coupling~\cite{Foerster1948,Forster2012,Roden2009,delRey2013} significantly affects transport properties. For multiple excitations we observe strong density-density correlations which persist at steady-state indicating effective interactions between excitations mediated by the non-trivial dissipation. This suggests a novel and physically realizable experimental route towards dissipative entanglement creation~\cite{Verstraete2009,Diehl2008,Weimer2010,Weimer2011,BhaktavatsalaRao2013} and realizing exotic pairing mechanisms or exotic quantum phases in the many-body regime~\cite{Diehl2010,Diehl2011}.

\acknowledgements{We acknowledge valuable discussions with Alexander Eisfeld, Michael Fleischhauer, Michael H\"oning and Martin Rabel. This work is supported in part by the Heidelberg Center for Quantum Dynamics, the European Union H2020 FET Proactive project RySQ (grant N. 640378), the 7th Framework Programme Initial Training Network (COHERENCE) and the Deutsche Forschungsgemeinschaft under WH141/1-1.}

\bibliography{spintransport}

\end{document}